\documentstyle[preprint,eqsecnum,aps]{revtex}
\def\BE{\begin{equation}}
\def\EE{\end{equation}}
\def\bx{{\bf x}}

\tightenlines

\begin{document}
\bibliographystyle{prsty}

\draft

\title{Numerical Study of a Lyapunov functional for the Complex
Ginzburg-Landau Equation}

\author{ R. Montagne
\footnote{on leave from Universidad de la Rep\'ublica (Uruguay).},
E. Hern\'andez-Garc\'\i a, and  M. San Miguel }
\address{Departament de F\'{\i}sica,
Universitat de les Illes Balears,  \\
and Institut Mediterrani d'Estudis Avan\c{c}ats, IMEDEA (CSIC-UIB)\\
E-07071 Palma de Mallorca (Spain) }

\date{\today}

\maketitle

\begin{abstract}

We numerically study in the one-dimensional case the validity of the
functional
calculated by Graham and  coworkers (R. Graham and T. Tel, Phys. Rev. A {\bf
42},
4661  (1990),  O. Descalzi and R. Graham, Z. Phys. B {\bf 93},  509  (1994))
as a
Lyapunov potential for the  Complex Ginzburg-Landau equation.  In non-chaotic
regions of parameter space the functional decreases  monotonically in time
towards the plane wave attractors, as expected for a Lyapunov functional,
provided that no phase  singularities are encountered. In the phase turbulence
region the potential  relaxes towards a value characteristic of the phase
turbulent attractor, and  the dynamics there approximately preserves a
constant value. There are however  very small but systematic deviations from
the
theoretical predictions, that increase when going deeper in the phase
turbulence region. In more disordered chaotic regimes characterized by the
presence of  phase singularities the functional is ill-defined and then not a
correct  Lyapunov potential.

\end{abstract}


\narrowtext
\section{INTRODUCTION}
\label{sec:Intro}

The Complex Ginzburg-Landau Equation (CGLE) is the amplitude equation
describing universal features of the dynamics of extended systems near a Hopf
bifurcation \cite{CrossHohenberg,hohenbergsaarloos}.
\begin{equation}
 \partial_{t} A = a A + ( D_r + {\it i} D_i ) \nabla^{2} A -
 (b_r + {\it i} b_i ) \mid A \mid^{2} A \ .
 \label{cgle}
\end{equation}
Examples of this situation include
binary fluid convection \cite{kolodner94}, transversally
extended lasers \cite{coullet} and chemical turbulence\cite{kuramoto81}. We
will considered here only the one-dimensional
case, $A=A(x,t)$, with $x \in [0,L] $. Suitable scaling of the complex
amplitude $ A $,
space, and time shows that for fixed sign of $a$ there are only
three independent parameters
in (\ref{cgle}) (with $D_r$ and $b_r>0$ that we assume henceforth). They can
be chosen to be $L$, $c_1 \equiv D_i/D_r$, and $c_2 \equiv b_i/b_r$.

The CGLE for $a>0$ displays a rich variety of complex spatio-temporal dynamical
regimes that have been recently classified in a phase diagram in the parameter
space $\{c_1,c_2\}$ \cite{chate1,chate2,chate3}. It is commonly stated that
such  nontrivial dynamical behavior, occurring also in  other nonequilibrium
systems, originates from the non-potential or  non-variational character of
the dynamics \cite{nonvariational}. This general statement needs to be
qualified because it involves some confusion in the terminology.  For example
the term ``non-variational" is often used meaning that there is no Lyapunov
functional for the dynamics. But Graham and co-workers,  in a series of papers
\cite{grahamtel90,grahamtel90a,grahamtel91,graham92,graham93}, have
shown that a Lyapunov functional does  exist for the CGLE, and they have
constructed it approximately in  a small-gradient approximation. The correct
statement for the CGLE is  that it is not a gradient flow. This means that
there is no
real functional of $A$ from which the right hand side of (\ref{cgle}) could be
obtained by functional derivation.

Part of the confusion associated with the qualification of
``nonvariational" dynamics comes from the idea that the dynamics of systems
having  non-trivial attractors, such as limit cycles or strange chaotic
attractors, can not be deduced from the minimization of a potential which plays
the
role of the free energy of equilibrium systems. However, such idea does not
preclude the existence of a Lyapunov functional for the dynamics. The Lyapunov
functional can have local minima which identify the attractors.
Once the system has reached an attractor which is not a fixed point,
dynamics can proceed on the attractor
due to ``nonvariational" contributions to the dynamical flow which do not
change
the value of the Lyapunov functional. This just means that the dynamical flow
is
not entirely determined once the Lyapunov functional is known. This situation
is very common and well known in the study of dynamical properties within the
framework of conventional statistical mechanics: The equilibrium free energy
of the system is a Lyapunov functional for the dynamics, but equilibrium
critical dynamics \cite{hohenberg78} usually involves contributions, such
as mode-mode coupling terms, which are not determined just by the free energy.
The fact that the dynamical evolution is not simply given by the minimization
of
the free energy is also true when studying the nonequilibrium  dynamics of a
phase transition in which the system evolves between an initial and a final
equilibrium state after, for example, a jump in temperature across the critical
point \cite{gunton83}.

A Lyapunov functional plays the role of a potential which is useful in
characterizing global properties of the dynamics, such as attractors,
relative or nonlinear stability of these attractors, etc. In fact, finding such
potentials is one of the long-sought goals of nonequilibrium physics
\cite{graham89,graham95}, the hope being that they should be instrumental in
the
characterization of  nonequilibrium phenomena through phase transitions
analogies. The use of powerful and very general methods based on these
analogies
has been advocated by a number of authors
\cite{ciliberto1,ciliberto2,chate1,chate2,chate3}.
In this context, it is a little surprising that the finding of a Lyapunov
functional for the CGLE \cite{grahamtel91,graham92,graham93} has not
received much attention in the literature. A possible reason for this is that
the
construction of nonequilibrium potentials has been historically associated
with the study of stochastic processes, in particular in the search of
stationary probability distributions for systems driven by random noise
\cite{graham89,graham95,tira3}. We want to make clear that the finding of the
Lyapunov
functional for the CGLE \cite{grahamtel91,graham92,graham93}, as well as the
whole approach and discussion the present paper is completely within a purely
deterministic framework and it does not rely on any noise considerations. A
second possible reason for the relative little attention paid to the Lyapunov
functional for the CGLE is the lack of any numerical check of the uncontrolled
approximations made on its derivation. The main purpose of this paper is
precisely to report such numerical check of the results of Graham and
collaborators, thus delimiting the range of validity of the approximations
involved. We also provide a characterization of the time evolution of the
Lyapunov functional in different regions of the phase diagram of the CGLE
\cite{chate1,chate2,chate3}, which illustrates the use of such potential.

Our main findings are that the expressions by Graham and coworkers behave
to a good approximation as a proper Lyapunov potential when phase
singularities (vanishing of the
modulus of $A$) are not present. This includes non-chaotic regimes as well
as states of phase turbulence. In this last case some small but
systematic discrepancies with the predictions are found. In the presence of
phase singularities the potential is ill-defined and then it is not a correct
Lyapunov functional.

The paper is organized as follows. For pedagogical  purposes, we first discuss
in
Sect. II a classification of dynamical flows in which notions like
relaxational
or potential flows are considered. The idea of a potential for the CGLE is
clearer
in this context. In Sect. III we review basic phenomenology of the CGLE  and
the
main analytical results for the Lyapunov functional of the CGLE.  Sections IV
and
V contain our numerical analyses. Section IV is devoted to the  Benjamin-Feir
stable regime of the CGLE and Sect. V to the Phase Turbulent regime. Our main
conclusions are summarized in Sect.VI.

\section{A classification of dynamical flows}
\label{pot}

\noindent In the following we review a classification of dynamical systems
that, although rather well established in other contexts
\cite{graham89,graham95}, it is
often overlooked in general discussions of deterministic spatio-temporal
dynamics.  Non-potential dynamical systems are often defined as those for which
there is  no Lyapunov potential. Unfortunately, this definition is also applied
to cases in which there is no {\sl known} Lyapunov potential. To be more
precise, let us consider dynamical systems of the general form
\BE   \partial_{t} A_i = V_i[A]
\label{ds}
\EE
where $A_i$ represents a set of, generally complex, dynamical variables
which are spatially dependent fields: $A_i=A_i(\bx,t)$.  $V_i[A]$ is a
functional of them.  The notation $A_i^*$ represents the
complex conjugate of $A_i$ and for simplicity we will keep the index $i$
implicit.
 Let us now split $V$  into two contributions:
 \BE
\label{split}  V[A]=G[A]+N[A] \ ,
\EE
\noindent where $G$, the {\sl relaxational} part, will have the form
\BE G[A]=-{\Gamma \over 2} {\delta F[A] \over \delta A^*} \  ,
\label{G}
\EE
\noindent with $F$ a real and scalar functional of $A$. $\Gamma$ is an
arbitrary hermitic and positive-definite operator (possibly depending
on $A$).  In the particular case of real variables there is no need of taking
the
complex conjugate, and hermitic operators reduce to symmetric ones. The
functional $N[A]$ in (\ref{split}) is the remaining part of
$V[A]$. The important point is that, if the splitting (\ref{split}) can
be done in such a way that the following orthogonality condition is
satisfied  (c.c. denotes the complex conjugate expression):
\BE
\int d\bx \left( {\delta F[A] \over \delta A(\bx)} N[A(\bx)] +
{\rm c.c.} \right)=0\  ,
\label{preHJ}
\EE
\noindent then the terms in $N$ neither increase nor decrease the value of
$F$,
which due to the terms in $G$ becomes a decreasing function of time:
\BE
{dF[A(\bx,t)] \over dt} \le 0 \ .
\label{decreasing}
\EE
If $F$ is bounded from below then it is a Lyapunov potential for the dynamics
(\ref{ds}). Equation (\ref{HJ}) with $N=V-G$, that is
\BE
\label{HJ}
\int d\bx \left( {\delta F[A] \over \delta A(\bx)}
\left( V[A(\bx)] + {\Gamma \over 2} {\delta F[A] \over \delta A^*(\bx)} \right)
+ {\rm c.c.} \right)=0\  ,
\EE
can be interpreted as an
equation  for the Lyapunov potential $F$ associated to a given dynamical system
(\ref{ds}). It has a  Hamilton-Jacobi structure. When dealing with systems
perturbed by random noise, $\Gamma$ is fixed by statistical requirements,
but in deterministic contexts such as the present paper, it can be arbitrarily
chosen in order to simplify (\ref{HJ}).

Solving (\ref{HJ}) is in general a difficult task, but a number of non-trivial
examples of the splitting  (\ref{split})-(\ref{decreasing}) exist in the
literature. Some of these examples correspond to solutions of (\ref{HJ})  found
in the search of potentials for dynamical systems
\cite{grahamtel91,grahamtel90,grahamtel90a}. Other examples just
correspond to a natural splitting of dissipative and non-dissipative
contributions in the dynamics of systems with well established  equilibrium
thermodynamics, as for example models of critical dynamics
\cite{hohenberg78} or the equations of nematodynamics in liquid crystals
\cite{sagues87}.

Once the notation above has been set-up, we can call  relaxational systems
those
for which there is a solution $F$ of (\ref{HJ}) such that $N=0$, that is all
the
terms in $V$ contribute to decrease $F$. Potential systems can be  defined as
those for which there is a nontrivial (i.e. a non-constant) solution $F$ to
(\ref{HJ}). In relaxational systems there is no long-time dynamics, since
there is no time evolution of $A$ once a minimum of $F$ is reached. On the
contrary,
for potential systems for which $N\neq 0$, the minima of $F$ define the
attractors of the dynamical flow, but once one of these attractors is reached,
nontrivial sustained dynamics might exist on the attractor. Such dynamics is
determined by $N$ and maintains a constant value for the functional $F$.

A possible more detailed classification of the dynamical flows is the
following:
\begin{itemize}
\item[1.-] Relaxational gradient flows: Those dynamical systems for which
$N=0$ with $\Gamma$ proportional to the identity operator. In this  case the
time
evolution of the system follows the lines of steepest descent of $F$. A well
known
example is  the so  called Fisher-Kolmogorov equation, also known as model A of
critical dynamics \cite{hohenberg78}, or (real) Ginzburg-Landau equation
for a real field $A(\bx,t)$:
\BE \label{modela}
\dot A = \alpha A + \gamma \nabla^{2} A - \beta \mid A \mid ^{2} A\  ,
\EE
\noindent where $\alpha$, $\gamma$  and $\beta$ are real coefficients. This
equation is of the form of Eqs. (\ref{ds})-(\ref{G}) with $N=0$, $\Gamma = 1$,
and $F = F_{GL}[A]$, the Ginzburg-Landau free energy:
\BE \label{GL}
F_{GL}[A] = \int d\bx \left( -{\alpha}\mid A \mid ^{2} +
{\gamma } \mid \nabla A \mid ^{2} + {\beta \over 2}  \mid A \mid ^{4}  \right)
\EE

\item[2.-] Relaxational non-gradient flows: Still $N=0$ but with $\Gamma$  not
proportional to the identity, so that the relaxation to the minimum of $F$
does
not follow the lines of steepest descent of $F$. The matrix operator $\Gamma$
might depend on $A$ or involve spatial derivatives. A well known example of
this
type is the Cahn-Hilliard equation of spinodal decomposition, or model B of
critical dynamics for a real variable $A$. \cite{hohenberg78}:
\BE \label{modelb}
\dot A = (- {1 \over 2}\nabla^2)
\left( -{\delta F_{GL}[A] \over \delta A} \right) \  ,
\EE
The symmetric and positive-definite operator $(- \nabla^2)$ has its origin in a
conservation law for $A$.

\item[3.-] Non-relaxational potential flows: $N$ does not vanish, but the
potential $F$, solution of (\ref{HJ}) exists and is non-trivial. Most models
used in equilibrium critical dynamics \cite{hohenberg78}  include
non-relaxational contributions, and therefore belong to this category.  A
particularly simple example is
\BE \label{ANLS}
\dot A = -(1+i) {\delta F_{GL}[A] \over \delta A^*} \ ,
\EE
\noindent where now $A$ is a complex field. Notice that we can not interpret
this
equation as being of type 1, because $(1+i)$ is not a hermitic operator,  but
still
$ F_{GL}$ is a Lyapunov functional for the dynamics. Equation (\ref{ANLS}) is a
special case of the Complex  Ginzburg- Landau Equation (CGLE), in which $V[A]$
is
the sum of a relaxational gradient flow and a nonlinear-Schr\"odinger-type
term $N[A]=-i {\delta F_{GL}[A] \over \delta A^*} \ $.

The general CGLE\cite{hohenbergsaarloos} is of the form (\ref{modela}) but
$A$ is complex and $\alpha$, $\gamma$  and $\beta$ are arbitrary complex
numbers. For
the special case in which $\frac{Re[\gamma]}{Im[ \gamma ]} =
\frac{Re[\beta]}{Im[\beta]}$, as for example in (\ref{ANLS}), the Lyapunov
functional for the CGLE is known exactly \cite{tel82}. Such choice of
parameters has important dynamical consequences\cite{tira2}.  Beyond such
special cases, the calculations by Graham and coworkers indicate
\cite{graham92,graham93} that the CGLE, a paradigm of complex
spatio-temporal dynamics, might be classified within this class of
non-relaxational potential flows because a solution of (\ref{HJ}) is found.
The difficulty is that the explicit form of the potential is, so far, only
known as
a uncontrolled small-gradient expansion.

 \item[4.-] Non-potential flows: Those for which the only solutions $F$ of
(\ref{HJ}) are the trivial ones (that is $F=$ constant). Hamiltonian systems as
for example the nonlinear Schr\"odinger equation are of this type.

\end{itemize}

\section{A LYAPUNOV FUNCTIONAL FOR THE CGLE}
\label{pheno}

It is well known that for $a<0$ the one dimensional CGLE (\ref{cgle}) has $A=0$
as a
stable solution, whereas for $a>0$ there are Travelling Wave (TW) solutions of
the form
\begin{equation}
A_k= A_{0}e^{i(k x +\omega t)+\varphi_0}
\label{tw}
\end{equation}
\noindent with $A_{0} = \sqrt{(a - D_r k^{2})/b_r}$, $|k|<\sqrt{a/b_r}$,
and $\omega = ( b_i a +D_- k^2)/b_r$.
We have introduced
\begin{equation}
D_{-}  \equiv  D_r b_i - D_i b_r  \ .
\label{lalinea}
\end{equation}
$\varphi_0$ is any arbitrary constant phase.

The linear stability of the homegeneus solution ( (\ref{tw}) with $k=0$ ) with
respect to long wavelength fluctuations divides the parameter space $ \{ c_1
,c_2 \}$ in two regions: the Benjamin-Feir (BF) stable and the  BF
unstable zone. This line is given by \cite{bf1,bf2}
\begin{equation}
 D_{+}  \equiv D_r b_r +D_i b_i = 0 ,
\label{bfline}
\end{equation}

In the BF unstable region ($D_+<0$) there are no stable TW solutions,
while in the BF stable
region ($D_+>0$)  TW's with a wavenumber $k<k_E$ are linearly stable. For
$k>k_E$, TW's
become unstable through the long wavelength instability known as the Eckhaus
instability \cite{eckhaus1,janiaud1}. The Eckhaus wavenumber $k_E$ is given by
\begin{equation}
k_E^2 = \frac{a b_r D_+}{D_r (3 D_+ b_r + 2 D_- b_i)}
\label{keckhaus}
\end{equation}

Recent numerical work for $a>0$ and $L$ large
\cite{chate1,chate2,chate3,egolf195}  has identified regions of the
parameter space displaying different kinds of regular and spatio-temporal
chaotic behavior (obtained at long times from  random initial conditions and
periodic boundary conditions), leading to a  ``phase diagram" for the CGLE. The
five different regions, each leading to a different asymptotic  phase, are
shown
in Fig.\ \ref{fig1}  as a function of the parameters $c_1$ and $c_2$ ($a>0$,
$L$
large).  Two of these regions are in the BF stable zone and the other three in
the BF
unstable one. One of the  main distinctions between the diferent asymptotic
phases is in the behavior of  the modulus of $A$ at long times. In some regions
it
never vanishes,  whereas in others it vanishes from time to time at different
points. A more  detailed description of the asymptotic behavior in the
different
regions is as follows:
\begin{enumerate}
\item Non-Chaotic region. The evolution here ends in one of the
Eckhaus-stable TW solutions for almost all the initial conditions.
\item Spatio-Temporal Intermittency region. Despite the fact that there
exist stable TW, the evolution from random initial conditions is not
attracted by them but by a chaotic attractor in which typical
configurations of the field $A$ consist of patches of TW interrupted by
turbulent bursts. The modulus of $A$ in such bursts typically touches zero
quite often.
\item Defect Turbulence. This is a strongly disordered phase in which
the modulus of $A$ has a
finite density of space-time zeros. In addition
the space and time correlation functions have a quasi-exponential
decay \cite{chate1,chate2}.
\item Phase Turbulence. This is a
weakly disordered phase in which $|A(x,t)|$ remains away from zero. The
temporal
correlations decay slower than exponentially \cite{chate1,chate2}.
\item Bi-Chaos region. Depending on the particular initial condition, the
system ends on attractors similar to the ones in regions 3, 4, or in a
new attractor in which the configurations of $A$ consists of
patches of phase and defect turbulence.
\end{enumerate}

An approximate Lyapunov functional for the CGLE was calculated by Graham and
collaborators \cite{graham92,graham93,descalzi}. Earlier attempts to find
a Lyapunov functional were based on polynomial
expansions\cite{tel82,graham75,walgraef82,walgraef83}, while more
recent and successful approaches focussed in solving the Hamilton-Jacobi
equation (\ref{HJ}) with $\Gamma=1$ in different ways. This was done first by a
minimization procedure involving an action
integral\cite{grahamtel90,grahamtel90a,grahamtel91}, and more recently
by a more direct expansion method \cite{graham92,graham93,descalzi}. This
last method provides also expressions in higher dimensions, but we will
restrict  here to the one-dimensional case.  In any case, the  solution
involves an
uncontrolled gradient expansion around space-independent solutions of the
CGLE. Such expansion obviously limits the validity of  the result to regions in
the phase diagram in which there are not strong gradients. Since  the expansion
was actually performed in polar coordinates, this excludes  the regions in
which
zeros in the modulus of $A$ are typical, since  the phase of $A$ becomes
singular
there. In particular  Spatio-temporal  intermittency regimes, Bi-chaos and
Defect Turbulence are out of the range of validity of Graham's expansion.  The
meaningfulness of the potential in the other regions of parameter space remains
still an open question because of the uncontrolled  small gradient
approximations used to calculate it, and calls for some numerical check.

In their solution of the Hamilton Jacobi equation, Graham and collaborators
find different branches of the Lyapunov functional with expressions valid  for
different values of the parameters. In particular  they identify the BF line
(\ref{bfline}) as separating two branches of  the solution to (\ref{HJ}).

The explicit expressions (obtained with $\Gamma=1$) are given in polar
coordinates:
\begin{equation}
A(x,t) = r(x,t)e^{i \varphi(x,t)}
\label{polar}
\end{equation}
In terms of the amplitude $r$, the phase $\varphi$, and their spatial
derivates (denoted as $r_x$, $\varphi_x$,$\varphi_{xx}$, etc.) the Lyapunov
functional per unit of length $\Phi \equiv F/L$
was found\cite{graham92,graham93}, for $a<0$:
\begin{equation}
\Phi   =  \int \biggl\{ b_{r} r^{4} - 2 a r^{2} + 2 \biggl[ D_{r} + \frac{D_{-}
 b_{i}
r^{4}} {3 ( a - b_{r} r^{2} )^{2} } \biggr] r_{x}^{2}  -  \frac{ 2 D_{-}
r^{3}} {3 ( a -
b_{r} r^{2} ) } r_{x} \varphi_{x} + 2 D_{r} r^{2} \varphi_{x}^{2}   \biggr\} dx
\label{anep0}
\end{equation}
\noindent We note that even in this relatively simple case $a<0$,  the result
for
$\Phi $ is only approximate  and its structure reveals a highly non-trivial
dynamics.

For  $a>0$, in  the BF stable region ($D_+>0$) the expression for $\Phi$
results:
\begin{eqnarray}
\Phi & = & \int \biggl\{ b_{r} r^{4} - 2 a r^{2} \nonumber \\
& + &\biggl[   (A_{1} r +
B_{1}/r^{2})r^{2}_{x} +   (A_{2} r + B_{2}/r) r_{x} \varphi_{x} +  2 (D_{r}
r^{2} -
D_{-} b_{i} a /     \mid b \mid^{2}b_{r} ) \varphi^{2}_{x} \biggr] \nonumber \\
& + &    \biggl[ \frac{D_{-} D_{r} b_{i}}{3 b_{r} \mid b \mid^{2}}
\varphi^{4}_{x} +
\biggl( - \frac{D_{-}^{2} a}{2 b^{4}_{r} r^{2}} -   \frac{D_{-}}{b_{r}^{2}} (
D_{-}/b_{r} + 2 D_{i} ) \ln r + C_{1} \biggr)  \varphi^{2}_{xx} + \frac{2 D_{-}
D_{r}}{3 b^{2}_{r} \mid b\mid^{2}} (b^{2}_{i}  - b^{2}_{r}
)\frac{\varphi^{3}_{x} r_{x}}{r} \nonumber \\
 & + & \frac{2 D_{-} D_{r} b_{i}}{3   b_{r}^{3} \mid b \mid^{2}} (b^{2}_{i} - 2
b^{2}_{r}) \frac{\varphi^{2}_{x}  r^{2}_{x}}{r^{2}} - \frac{4
D_{r}b_{i}D_{-}}{3 b^{3}_{r} r} \biggl( 1 + \frac{ \ln(b_{r}   r^{2}/a) }{1 -
b_{r} r^{2}/a } \biggr) r_{x} \varphi_{x} \varphi_{xx} \biggr] \biggr\} dx
\label{anep1}
\end{eqnarray}
where
\begin{eqnarray}
 A_{1} &=& 2 (D_{r} + b_{i}D_{-}/3b_{r}^{2}) ,\nonumber \\
 A_{2} &=& 2 D_{-}/ b_{r} ,\nonumber \\
 B_{1 }&=& \frac{2 D_{-} b_{i} a}{3 b_{r}^{3} \mid b \mid^{2}} (2 b^{2}_{r}
 - b^{2}_{i} ) ,\nonumber \\
 B_{2} &=& \frac{2 D_{-}  a}{  b_{r}^{2} \mid b \mid^{2}} ( b^{2}_{r}
 - b^{2}_{i} ) ,\nonumber \\
\end{eqnarray}
Clearly, $\Phi$ is ill-defined when $r=0$.

By writing-out the Euler-Lagrange equations associated to the minimization  of
$\Phi$ the TW solutions (\ref{tw}) are identified as local extrema of $\Phi$.
Since they occur in families parametrized by the arbitrary phase  $\varphi_0$,
the minima associated to the TW of a given $k$ are not isolated  points but lay
on a
one-dimensional closed manifold. The non-variational part  of the  dynamics
($N$ in (\ref{split})) can be explicitly written-down by substracting  $G=-{1
\over 2} {\delta F \over \delta A^*}$ with $F=L\Phi$  to the right-hand-side of
(\ref{cgle}). It is seen to produce, when evaluated on the manifold  of minima
of
$\Phi$ with a given $k$, constant motion along it. This produces   the periodic
time dependence in (\ref{tw}) and identify the  TW attractors as limit cycles.

The value of  $k$ for which the corresponding extrema change character from
local
minima  to saddle points is precisely the Eckhaus wavenumber $k_E$.  It is
remarkable that, although expression (\ref{anep1}) was obtained in a  gradient
expansion around the homogeneous TW, their minima identify exactly  all the
TW's
of equation (\ref{cgle}), and their frequencies and points  of instability are
also exactly reproduced. This gives confidence on the  validity of Graham's
approximations. It should be stressed however that  they are not exact and can
lead to unphysical consequences. For instance,   the value of the potential
$\Phi$ evaluated on a TW of wavenumber $k$ ($|k|<\sqrt{a/b_r}$) is
\cite{grahamtel91}
\begin{equation}
\Phi_{k} \equiv \Phi[A_k] = \frac{2 D_{+} a}{\mid b \mid^2} k^{2}  \left( 1 -
\frac{k^{2}}{6 k_{E}^{2}} \right) + \Phi_{k=0}
\label{fisuk}
\end{equation}
\noindent where $\Phi_{k=0}=  - a^{2}/b_r$. For a range of parameter values
this
expression gives  mathematical sense to the intuitive fact that the  closer to
zero is $k$  the more stable is the associated TW (because its potential is
lower).
But for some parameter values the minimal potential corresponds to large
wavenumbers close to $\pm \sqrt{a/b_r}$. This is counterintuitive and calls
for some numerical test. The test will be described below and it will  be shown
that
the wavenumbers close to $\pm \sqrt{a/b_r}$ are out of the range  of validity
of
the small gradient approximations leading to (\ref{anep1}).

We already mentioned in the previous section that the Lyapunov functional for
the CGLE is exactly known for special values of the   parameters
\cite{grahamtel91,graham92,tira2}.  This happens for $D_{-}\equiv D_r b_i -
D_i b_r = 0$, which lies in  the BF-stable region as indicated in Fig.\
\ref{fig1} .
In this case it is clear that (\ref{cgle}) can be written as
\begin{equation}
\dot A = -  {1 \over 2}{\delta F_{GL}[A] \over \delta A^*} +  i b_i \left(-\mid
A
\mid^2 + \frac{D_r}{b_r}\nabla^2 \right) A \ ,
\label{special}
\end{equation}
\noindent where $ F_{GL}[A]$ is (\ref{GL}) for complex $A$ and with
$\alpha=2a$, $ \beta = 2 b_r$, and $\gamma = 2 D_r$. It is readily shown  that
the
term  proportional to $b_i$ is orthogonal to the gradient part, so that
$F_{GL}$
is an exact solution of (\ref{HJ}) for these values of the parameters, and
(\ref{special}) is a relaxational non-gradient flow (see classification in
section \ref{pot}). It is seen that the approximate expressions  (\ref{anep0})
and (\ref{anep1}) greatly simplify when $D_-=0$ leading both  to the same
expression:
\begin{equation}
 L \Phi =   \int \biggl\{- 2 a r^{2}  + b_{r} r^{4} +    2 D_{r} r  r^{2}_{x}
+ 2 D_{r} r^{2} \varphi^{2}_{x}  \biggr\} dx
\label{anepsp}
\end{equation}
When expressed in terms of $A$ and $A^*$ it reproduces $F_{GL}$ in
(\ref{special}). Thus the gradient expansion turns out to be exact on the  line
$D_-=0$.

In the Benjamin-Feir unstable region ($a>0, D_+<0$) the gradient expansion for
$\Phi$ becomes\cite{graham93,descalzi}:
\begin{eqnarray}
\Phi & = & \int \biggl\{ b_{r} r^{4} - 2 a r^{2} +\biggl[   (A_{1} r +
\tilde{B_{1}}/r^{2})r^{2}_{x} +   (A_{2} r + \tilde{B_{2}}/r) r_{x} \varphi_{x}
+
2 D_{r}( r^{2} -\frac{a}{b_{r}})  \varphi^{2}_{x} \biggr]
\nonumber \\ & + &
\biggl[ \frac{D_{r}^{2} }{ b_{r} }  \varphi^{4}_{x} + \biggl(  \frac{b_{r}}{2
a^{2} r^{2}} \biggl( \frac{\tilde{B_{2}}^{2}}{4} + \frac{4 D_{r}^{2}
a^{2}}{b_{r}^{2}} \biggr) \bigl(r^{2} - \frac{a}{b_{r}} \bigr)  -
\frac{A_{2}}{2 b_{r}} \bigl(  \frac{A_{2}}{4} + D_{i} \bigr) \ln \bigl(
\frac{r^{2} b_{r}}{a} \bigr) + \frac{D_{i}^{2}}{b_{r}} \biggr) \varphi^{2}_{xx}
\nonumber \\ & - &
\frac{4 D_{r} b_{i} D_{-}}{b_{r}^{3} r} \biggl( 1+ \frac{D_{r}\mid b\mid^{2} +
2 b_{r} D_{+}}{b_{i} D_{-} (1 - \frac{b_{r} r^{2}}{a})} \ln \bigl( \frac{r^{2}
b_{r}}{a} \bigr) \biggr) \varphi_{x} r_{x}\varphi_{xx}
\nonumber \\ & + &
\frac{2 D_{r}}{3 b_{r}^{2} r} ( 5 b_{i} D_{r} + D_{i} b_{r} ) \varphi^{3}_{x}
r_{x} +
\frac{2 b_{i}D_{r}}{3 b_{r}^{3} r^{2}} ( 7 b_{i}D_{r}+ D_{i} b_{r})
\varphi^{2}_{x} r^{2}_{x}   \biggr] \biggr\} dx
\label{anep3}
\end{eqnarray}
where, in addition to the previous definitions
\begin{eqnarray}
\tilde{B_{1 }}&=& \frac{2 D_{-} b_{i} a}{3 b_{r}^{3} \mid b \mid^{2}} (2
b^{2}_{r}
 - b^{2}_{i} ) ,\nonumber \\
\tilde{B_{2}} &=& - \frac{2  a}{b_{r}^{2} } (D_{r} b_{i} + D_{i} b_{r} ) ,
\nonumber \\
\end{eqnarray}
It was noted before that this expression can be adequate, at most, for the
Phase Turbulent regime, since in the other BF unstable regimes $|A|$ vanishes
at some points and instants, so that (\ref{anep3}) is ill-defined.

The long time dynamics occurs in the attractor defined by the minima of
$\Phi$. The Euler-Lagrange equations   associated to the
minimization of (\ref{anep3}) lead to a relationship between
amplitude and phase of $A$ which implies the well known adiabatic following of
the amplitude to the phase dynamics commonly used to describe the phase
turbulence regime by a nonlinear phase equation. The explicit form of this
relationship is
\begin{eqnarray}
r^2 & = & \frac{a}{br} - \frac{D_r}{b_r} (\nabla \varphi)^2 - \frac{D_i}{b_r}
\nabla^2 \varphi +  \frac{b_i D_i^2}{2 a b_r^2} \nabla^4 \varphi  +   2
\frac{D_r
D_i b_i}{a b_r^2} \nabla \varphi \nabla^3 \varphi \nonumber \\
 & +& 2 \frac{b_i D_r^2}{a b_r^2}  \nabla \varphi \nabla \nabla^2 \varphi
+   \left[
\frac{D_r D_i b_i}{a b_r^2} - \frac{ \mid D \mid^2 }{a b_r} \right] (\nabla^2
\varphi)^2
\label{radiab}
\end{eqnarray}

It defines the attractor characterizing the phase turbulent
regime. Dynamics in this  attractor follows from the nonrelaxational part
$N$ in (\ref{split}). When (\ref{radiab}) is imposed in such nonrelaxational
part of the dynamics the generalized Kuramoto-Shivashinsky equation containing
terms up to fourth order in the gradients \cite{sakaguchi2}
is obtained \cite{graham93,descalzi}.

We finally note that in the phase turbulent
regime the Lyapunov functional $\Phi$ gives the same value
\cite{graham93,descalzi} when evaluated for any
configuration satisfying (\ref{radiab}), at least within the small gradient
approximation. This corresponds to the evolution on a
chaotic attractor (associated to the Kuramoto-Sivashinsky dynamics coming from
$N$) which is itself embedded in a region of constant $\Phi$ (the potential
plateau \cite{graham95}). This plateau consists of the functional
minima of $\Phi$ (\ref{radiab}). All the (unstable) TW are also contained
in the same plateau, since they satisfy (\ref{radiab}).

\section{NUMERICAL STUDIES OF THE LYAPUNOV FUNCTIONAL IN THE BENJAMIN-FEIR
STABLE REGIME}
\label{results}

We numerically investigate the validity of $\Phi[A]$ in (\ref{anep0}),
(\ref{anep1}), and (\ref{anep3}) as an approximate Lyapunov functional for
the CGLE. When evaluated on solutions $A(x,t)$ of (\ref{cgle})  it should
behave
as a  monotonously decreasing function of time, until $A(x,t)$ reaches  the
asymptotic attractor. After then, $\Phi$ should maintain in time  a constant
value characteristic of the particular attractor.

All the results reported here were obtained using a pseudo-spectral code with
periodic boundary conditions and second-order accuracy in time. Spatial
resolution was typically 512 modes, with runs of up to 4096 modes to  confirm
the
results. Time step was typically $\Delta t = .1$ except when differently stated
in the figure  captions. Since very small effects have been explored, care has
been  taken of confirming the invariance of the results with decreasing time
step
and increasing number of modes.  System size was always taken as $L=512$, and
always $D_r =1$ and $b_i=-1$,  so that $c_1=D_i$ and $c_2=-1/b_r$. When a
random
noise of amplitude  $\epsilon$ is said to be  used as or added to an initial
condition it means that  a set of uncorrelated Gaussian numbers of zero mean
and
variance  $\epsilon^2$ was generated, one for each collocation point in the
numerical lattice.

\subsection{Negative $a$}
\label{amenor0}
The uniform state $A =0$ is stable for $a<0$. We start our numerical
simulation
with a plane wave $A=A_0 e^{ikx}$ of arbitrary wavenumber $k=0.295$ and
arbitrary amplitude $A_0=1$ (note that the TW's (\ref{tw}) do not exist  for
$a<0$), and calculate $\Phi$ for the evolving configurations.  In order to have
relevant nonlinear effects during the relaxation towards $A=0$ we have chosen
a
small value for the coefficient of the linear term  ($a= -0.01$). The remaining
parameters were $D_i=1$ and $b_r=1.25$  ($c_1=1$, $c_2=-0.8$).  Despite the
presence of non-relaxational terms in (\ref{cgle}),  $\Phi$ decreases
monotonously (see Fig.\ \ref{fig2}) to the final value  $\Phi( t = \infty ) =
\Phi[A=0] = 0$ confirming its adequacy as a Lyapunov potential.

\subsection{Positive $a$. Benjamin-Feir stable regime}
\label{bajobf}

We take in this section always $a=1$. Non-chaotic (TW) states and
Spatio-Temporal Intermittency are the two phases  found below the BF line in
Fig.\ \ref{fig1}. We first perform several  numerical experiments in the
non-chaotic region:

A first important case is the one on the line $D_{-} = 0$, for which
(\ref{anepsp})
is an exact Lyapunov functional $F_{GL}$. We take  $D_i=-1$ and $b_r=1$
($c_{1}=c_{2} = -1$), on  the $D_-=0$ line, and compute the evolution of
$\Phi=\frac{ F_{GL}}{L}$  along  a solution of (\ref{cgle}), taking as initial
condition for $A$ a Gaussian  noise of amplitude $\epsilon = 0.01$. Despite of
the
strong phase gradients  present   specially in the initial stages of the
evolution, and of the presence of  non-relaxational  terms, $\Phi$ decays
monotonously in time (Fig.\ \ref{fig3}). The system  evolved  towards a TW
attractor of wavenumber $k = 0.0245$. The value of $\Phi$ in  such  state is,
from
Eq. (\ref{fisuk}), $\Phi_{k=0.0245}=-0.998796$. It is  important to notice
that our numerical solution for $A$ and numerical  evaluation of the
derivatives
in $\Phi$ reproduce this value within a  $0.3 \%$ in the last
time showed in Fig.\ ref{fig3}, and continues to  approach the  theoretical
value
for the asymptotic attractor at longer times\footnote{If a smaller time step is
used greater
accuracy is obtained. For example, if the time step is reduced to $0.05$ the
value
of $\Phi$ is reproduced within $10^{-7} \%$. But this takes quite a long
computing time.}.

We continue testing the Lyapunov functional for $D_i=1$, $b_r=1.25$
($c_{1}=1$ ,$c_{2} = -0.8$.  This is still in the non-chaotic region but, since
$D_-\neq0$,    $\Phi$ is not expected to be exact, but only a small gradient
approximation.  We check now the relaxation back to an stable state after a
small
perturbation.  As initial condition we slightly perturb a TW of Eckhaus-stable
wavenumber ($k=0.13<k_{E}$) by adding random noise  of amplitude $\epsilon
=0.09$. $\Phi$ decays monotonously  (Fig.\ \ref{fig4}) from its  perturbed
value to the value $\Phi_{k=0.13}=-0.796632$ as the perturbation  is being
washed out, as expected for a good Lyapunov functional.

A more demanding situation was investigated for $D_i= -1$ and $b_r = 0.5$
(again in the non-chaotic region, $c_{1} = -1$ and $c_{2} = -2$,
and $D_-\neq 0$). Two TW of different
wavenumbers ($k_1 = 0.4 , k_2 = 0.08$, both Eckhaus-stable) were joined
and the resulting state
(see inset in Fig.\ \ref{fig5}) was used as initial condition. The TW of
smaller
wavenumber advances into the other, in agreement with the idea that it is
nonlinearly more stable since it gives a smaller value to the potential.
As the
difference between the two frequencies is large the speed at which one wave
advances onto the other is quite large. The interface between the two
TW's contains initially a discontinuity in the gradient of the phase which
is washed out in a few integration steps. An important observation is that
during the whole process the modulus of $A(x,t)$ never vanishes and then
the winding number, defined as
\BE
\label{winding}
\nu \equiv\int_0^L \nabla \varphi dx
\EE
remains constant ($\nu=20$) (with periodic boundary conditions $\nu$ is
constant except at the instants in which the phase becomes singular, that is
when
$r=0$). After the TW with the smallest wavenumber completely replaced the
other, still a phase diffusion  process in which the wave adjusts its local
wavenumber to the global  winding number occurs. The state (limit cycle)
finally
reached is a TW of  $k = 2\pi\nu/L=0.245$. Despite of the complicated and
non-relaxational  processes occurring $\Phi$   behaves as a good Lyapunov
functional monotonously decreasing from the  value $\Phi(t=0) = -1.825$
corresponding to the two-wave configuration  to the value $\Phi = -1.863$ of
the
final attractor (Fig.\ ref{fig5}).  It would be interesting, as happens in some
relaxational models  \cite{chan}, finding some relationship between the speed
of propagation of the more stable wave onto the less stable one and the
difference
in $\Phi$ between the two states.

The good behavior of $\Phi$ will be obviously lost if the  field $A(x,t)$
vanishes
somewhere during the evolution. As the next  numerical experiment (for $D_i=1$
and $b_r=1.25$, that is $c_{1} =  1$,  $c_{2} =-0.8$) we used as initial
condition a
small ($\epsilon=0.01$)  random Gaussian noise. The system was left to evolve
towards its asymptotic  state (a TW).  Fig.\ \ref{fig6} shows that after a
transient $\Phi$ monotonously decreases.  During  the initial transient it
widely fluctuates, increasing and decreasing  and loosing then its validity as
a
Lyapunov functional. This  incorrect behavior occurs because during the
initial stages $A(x,t)$ is  small and often vanishes, changing $\nu$. When $A$
(and then $r$) vanishes  the phase and (\ref{anep1}) are ill-defined and out of
the range of validity  of a small gradient approximation. Note the contrast
with
the case $D_-=0$ in  which the potential is exact and well behaved  even when
$\nu$
is strongly changing. The particular  values of the maxima and minima during
the
transient in which $\nu$ is changing   depend on the spatial and temporal
discretization, since it is clear from  (\ref{anep1}) that $\Phi$ is
ill-defined or divergent when $r$ vanishes.  Note that this incorrect behavior
of $\Phi$ for $D_-\neq 0$ is not a problem  for the existence of a Lyapunov
functional, but comes rather from the  limited validity of the hypothesis  used
for its approximate construction. Nevertheless, as soon as the strong
gradients disappear $\Phi$ relaxes monotonously to the value
$\Phi=-0.79997$,  corresponding to the final state, a TW of wavenumber $k =
-0.0123$.

As another test in the non-chaotic region, for $D_i=-1$ and $b_r=0.5$  ($c_{1}
=
-1$, $c_{2} = -2$)   we use as initial condition an Eckhaus-unstable TW
($k=0.54>k_{E}=0.48$)  slightly perturbed by noise. The system evolves to an
Eckhaus-stable TW  ($k = 0.31$) by decreasing its winding number (initially
$\nu
= 44$  and finally $\nu = 26$). Fig.\ \ref{fig7} shows the evolution of  $\Phi$
from
its initial value $\Phi(0) = -1.485$ the final one $\Phi = -1.77$.  Although
there
is a monotonously decreasing baseline, sharp peaks are  observed corresponding
to the vanishing of $r$ associated with the  changes in $\nu$. When $\nu$
finally
stops changing, so that $A$ is close enough to the final TW,  $\Phi$ relaxes
monotonously as in Fig.\ \ref{fig4}.

It was explained in Sect. \ref{pheno} that there are parameter ranges in which
$\Phi$ is smaller near the boundaries for existence of TW, that is near
$k =\pm \sqrt{a/b_r}$, than for the homogeneous TW: $k=0$. This happens for
example for $D_i=1$, $b_r=1.25$ ($c_{1} = 1$, $c_{2} = -0.8$). The
corresponding function $\Phi_k$ is shown
in Fig.\ \ref{fig8}. If this prediction is true, and if $\Phi$ is a
correct
Lyapunov functional, evolution starting with one of these extreme and
Eckhaus-unstable TW would not lead to any final TW, since this would increase
the value of the Lyapunov functional. This would imply the
existence for this value of the parameters of an attractor different from
the TW's perhaps related to the Spatio-Temporal Intermittency phenomenon.
We use as initial condition at the parameter values of Fig.\ \ref{fig8}
an unstable TW of wavenumber $k = 0.64$ ($\Phi \approx -0.81$), slightly
perturbed by noise. From Fig.\ \ref{fig8}, the system should evolve to a state
with a value of $\Phi$ value even lower than that. What really happens
can be seen in Fig.\ \ref{fig9}. The system changes its winding number from
the initial value $\nu=52$,
a process during which $\Phi$ widely fluctuates and is not a correct
Lyapunov functional, and ends-up in a state
of $\nu =  5$, with a value of $\Phi$ larger than the initial one. After
this the system relaxes to the associated stable TW of
$k=2\pi\nu/L=0.061 < k_{E}= 0.23$. As clearly stated by Graham and coworkers,
the expressions for the potential are only valid for small gradients. Since
$k$ is a phase gradient, results such as Fig.\ \ref{fig8}  can only be trusted
for $k$ small enough.

Finally, we show the behavior of $\Phi$ in the
Spatio-Temporal Intermittency regime. Since $\nu$ is constantly changing
in this regime it is clear that (\ref{anep1}) will not be a good Lyapunov
functional and this simulation
is included only for completeness. We take $D_i=0$ and $b_r=0.5$
($c_1=0$, $c_2=-2$) and choose as initial condition a
TW with $k= 0.44>k_{E}= 0.30$ ($\Phi=-1.89814$), with a small amount of noise
added. The TW decreases its winding number and the system reaches soon
the disordered regime called Spatio-Temporal Intermittency. Fig.\ \ref{fig10}
shows
that the time evolution of $\Phi$ is plagued with divergences, reflecting
the fact that $\nu$ is constantly changing (see inset). It is interesting
to observe however that during the initial escape from the unstable TW
$\Phi$ shows a decreasing tendency, and that its average value in the
chaotic regime, excluding the divergences, seems smaller than the initial one.

\section{NUMERICAL STUDIES OF THE LYAPUNOV FUNCTIONAL IN THE PHASE TURBULENCE
REGIME}
\label{pt}

The Phase Turbulence regime is characterized by the
absence of phase singularities (thus $\nu$ is constant). This
distinguishes it as the only chaotic regime for which $\Phi$ would be
well-defined. Graham and
co-workers\cite{graham93,descalzi} derived especially for this region an
expression proposed as Lyapunov functional in the small gradient
approximation (\ref{anep3}).

We recall that the calculations in \cite{graham93,descalzi} predict that
the phase turbulent attractor lies on a potential plateau, consisting
of all the complex functions satisfying (\ref{radiab}), in which all the
unstable TW cycles are also embedded. The value of the potential on such
plateau can be easily calculated by substituting in (\ref{anep3}) an arbitrary
TW, and the result is
\BE
\label{plateau}
\Phi_{pl}=-{a^2 \over b_r} \ .
\EE
We note that this value does not depend on $D_i$ nor $D_r$ and then it
is independent of $c_1$, the vertical position in the diagram of
Fig.\ \ref{fig1}, within the phase turbulence region.

In this section we take also $a=1$. We perform different simulations
for $D_i=1.75$ and $b_r=1.25$ ($c_1=1.75$, $c_2=-0.8$).
In the first one, we start the evolution with the homogeneous oscillation
solution (TW of $k=0$). This
solution is linearly unstable, but since no perturbation is added, the
system does not escape from it. The potential value predicted by
(\ref{plateau}) is $\Phi_{pl}=-0.8$. This value is reproduced by the
numerical simulation up to the sixth significant figure for all times
(Fig.\ \ref{fig11}, solid line). This agreement,
and the fact that the unstable TW is maintained, gives confidence in
our numerical procedure.

In a second simulation, a smooth perturbation (of the form $\mu e^{iqx}$ with
$q=0.049$ and $\mu=0.09$) is
added to the unstable TW and the result used as initial condition. This choice
of perturbation was taken to remain as much as possible within the range of
validity of the small gradient hypothesis. After a transient the perturbation
grows and the TW is
replaced by the phase turbulence state (the winding number remains fixed to
$0$).
The corresponding evolution of $\Phi$
is shown in Fig.\ \ref{fig11}  (long-dashed line). The value of the potential
increases
from $\Phi_{pl}$ to a higher value, and then irregularly oscillates around it.
Both the departure and the fluctuation are very small, of the order of
$10^{-4}$ times the value of $\Phi$. Simulations with higher precisions
confirm that these small discrepancies from the theoretical predictions are
not an artifact of our
numerics, but should be attributed to the terms with higher gradients
which are not included in (\ref{anep3}). As a conclusion, the prediction
that the phase turbulence dynamics, driven by non-relaxational terms,
maintains constant $\Phi$ in a value equal to the one for TW is confirmed
within a great accuracy.

It is interesting however to study how systematic
are the small deviations from the theory. To this end we repeat the
launching of the TW with a small perturbation for several values of
$D_i=c_1$,
for the same value of $b_r$ as before. The prediction is that $\Phi$
should be independent of $c_1$. The inset in Fig.\ \ref{fig11}  shows that
the theoretical value $\Phi_{pl}=-0.8$ is attained near the BF line, and
that as $c_1$ is increased away from the BF line there are very small but
systematic discrepancies. The values shown for the potential are time averages
of its instantaneous values, and the error bars denote the standard deviation
of the fluctuations around the average.

Again for $c_1=1.75$, $c_2=-0.8$, we perform another simulation
(Fig.\ \ref{fig11}, short-dashed line) consisting in
starting the system in a random Gaussan  noise configuration, of
amplitude $0.01$, and letting it to evolve towards the phase turbulence
attractor.
As in other cases, there is a transient in which $\Phi$ is ill-defined since
the winding number is constantly changing. After this $\Phi$ decreases. This
decreasing is not monotonous but presents small fluctuations around a
decreasing trend. The decreasing finally stops and $\Phi$ remains oscillating
around approximately the same value as obtained from the perturbed TW initial
condition. The final state has $\nu=-1$, so that in fact the attractor reached
is different from the one in the previous runs ($\nu=0$) but the difference
is the smallest possible and the difference in value of the associated
potentials can not be distinguished within the fluctuations of
Fig.\ \ref{fig11}.
These observations confirm the idea of a potential which decreases as
the system advances towards an attractor, and remains constant there, but
at variance with the cases in the non-chaotic region here the decreasing
is not perfectly monotonous, and the final value is only approximately
constant.

Since the small discrepancies with the theory increase far from the BF line,
and
since it is known that condition (\ref{radiab}) can be obtained from  an
adiabatic-following of the modulus to the phase that losses accuracy  far from
the BF line, one is lead to consider the role of adiabatic following  on the
validity of $\Phi$ as a potential. To this end we evaluated $\Phi$  along
trajectories $A(x,t)$ constructed with the phase obtained from  solutions of
(\ref{cgle}), but with modulus replaced by (\ref{radiab}),  so enforcing the
adiabatic following of the modulus to the phase. No  significant improvement
was
obtained with respect to the cases in which  the adiabatic following was not
enforced since that, in fact, adiabatic following was quite welll accomplished
by the solution of (\ref{cgle}). Then it is not the fact that  the solutions of
(\ref{cgle}) do not fulfill (\ref{radiab}) exactly,  but the absence of higher
gradient terms in both (\ref{radiab}) and  (\ref{anep3}) the responsible for
the small failures in the behavior of  $\Phi$.

Finally, it is interesting to show that the Lyapunov potential $\Phi$
can be used as a diagnostic tool
for detecting changes in behavior that would be difficult to monitor by
observing the complete state of the system. For example the time
at which the phase turbulence attractor is reached can
be readily identified from the time-behavior of $\Phi$ in Fig.\ \ref{fig11}.
More interestingly it can be used to detect the escape from metastable states.
For example, Fig.\ \ref{fig12}  shows $\Phi$ for evolution from a Gaussian
noise initial condition ($\epsilon=0.01$). $D_i=2$ and $b_r=1.25$ ($c_1=2$,
$c_2=-0.8$). The system reaches first a long lived state with $\nu=2$
not too different from the usual phase turbulent state of $\nu=2$. After
a long time however the system leaves this metastable state and approaches
a more ordered state that can be described \cite{montagne3} as phase turbulent
fluctuations around quasiperiodic configurations related to those of
\cite{janiaud1}. More details about this state will be described elsewhere
\cite{montagne3}. What is of interest here is that from Fig.\ \ref{fig12}
one can easily identify the changes between the different dynamical
regimes. In particular the decrease in the fluctuations of $\Phi$ near
$t \approx 1000$ identifies the jump from the first to the second
turbulence regimes.

\section{CONCLUSIONS AND OUTLOOK}
\label{conclusiones}

The validity of the expressions for the Lyapunov
functional of the CGLE found by Graham and coworkers has been numerically
tested. The most important
limitation is that they were explicitly constructed in a
approximation
limited to small gradients of modulus and phase. This precludes its
use for evolution on attractors such that zeros of $r$ and thus
phase singularities appear
(defect turbulence, bi-chaos, spatio-temporal intermittency).
The same problem applies to transient states of evolution towards more regular
attractors, if phase singularities appear in this transient (for instance
decay of an Eckhaus unstable TW, evolution from random states close to $A=0$,
etc.). A major step forward would be the calculation of the Lyapunov
potential for small gradients of the real and imaginary components of $A$,
which would be a well behaved expansion despite the presence
of phase singularities.

Apart from this, if changes in winding number are avoided, expressions
(\ref{anep0}), (\ref{anep1}), and (\ref{anep3}) display the correct properties
of a Lyapunov functional: minima on stable attractors, where non-relaxational
dynamics maintains it in a constant value, and decreasing value during
approach to the attractor. These properties are completely satisfied in
the non-chaotic region of parameter space, even in complex situations
such as TW competition, as long as large gradients do not appear.
It is remarkable that, although the potential is constructed trough an
expansion around the $k=0$ TW, its minima identify exactly the remaining
TW, its stability, and the non-relaxational terms calculated by substracting
the potential terms to (\ref{cgle}) give exactly their frequencies.
In the phase turbulence regime, however, there are small discrepancies with
respect to the theoretical predictions: lack of monotonicity in the approach
to the attractor, small fluctuations around the asymptotic value, and
small discrepancy between the values of the potential of TW's and of
turbulent configurations, that were predicted to be equal. All
these deviations are very small but systematic, and grow as we go deeper
in the phase turbulence regime.  They can be fixed in principle by
calculating more terms in the gradient expansion.

In addition in order to clarify  the conceptual status of non-relaxational and
non-potential dynamical systems one can ask about the utility of having
approximate expressions for the Lyapunov functional of the CGLE. Several
applications have been already developped for the case in which (\ref{cgle})
is perturbed with random noise. In particular the stationary probability
distribution is directly related to $\Phi$, and in addition barriers and
escape times from metastable TW have been calculated
\cite{grahamtel91,descalzi}. In the absence of random noise, $\Phi$ should
be still useful in stating the nonlinear stability of the
different attractors. In practice however there will be limitations in
the validity of the predictions, since $\Phi$ has been constructed in an
expansion which is safe only near one particular attractor
(the homogeneous TW).

Once known $\Phi$, powerful statistical mechanics techniques  (mean field,
renormalization group, etc. ) can in principle  be applied to it to obtain
information on the static properties of  the CGLE (the dynamical properties, as
time-correlation functions, would  depend also on the non-relaxational terms
$N$, as in critical dynamics  \cite{hohenberg78}). Zero-temperature Monte
Carlo  methods can also be applied to sample the phase turbulent attractors, as
 an
alternative to following the dynamical evolution on it. All those promising
developments will have to face first with the complexity of  Eqs.
(\ref{anep0}),
(\ref{anep1}), and (\ref{anep3}). Another use of  Lyapunov potentials (the one
most used in equilibrium thermodynamics)  is the identification of attractors
by minimization instead of by solving  the dynamical equations. In the case of
the
TW attractors, solving  the Euler-Lagrange equations for the minimization of
$\Phi$ is in fact more  complex than solving directly the CGLE with a TW
ansatz. But
the limit cycle character of the attractors, and their specific form, is
derived, not guessed as when substituting the TW ansatz. For the case  of
chaotic
attractors (as in the phase turbulence regime) minimization  of potentials can
provide a step towards the construction of  inertial manifolds. In this
respect
it should be useful considering the relationships between the Lyapunov
potential of Graham and coworkers and other objects based on functional  norms
used also to characterize chaotic attractors \cite{doering1,doering2}.

\section{ Acknowledgments}
We acknowledge very helpful discussions on the subject of this paper with R.
Graham.  We also acknowledge helpful inputs of E. Tirapegui and R. Toral on the
general ideas of nonequilibrium potentials. RM and EHG acknowledge financial
support from DGYCIT (Spain)
Project PB92-0046. R.M. also acknowledges partial support from the Programa de
Desarrollo de
las Ciencias B\'asicas (PEDECIBA, Uruguay), the Consejo Nacional de
Investigaciones Cient\'\i ficas Y T\'ecnicas (CONICYT, Uruguay) and the
Programa de Cooperaci\'on con Iberoam\'erica (ICI, Spain).


\newpage

\begin{figure}
\caption{ Regions of the parameter $[c_{1}\!-\!c_{2}]$-space ($a= 1$)  for the
$d=1$ CGLE displaying different kinds of regular and  chaotic behavior. Two
analytically obtained lines,the Benjamin-Feir line (B-F line) and the $D_{-}$
line,  are also shown.  }
\label{fig1}
\end{figure}

\begin{figure}
 \caption{Relaxation to the simple attractor for $a<0$. The parameter values
are $a =-0.01 , c_{1} = 1 \mbox{ and } c_{2} = - 0.8$ .  The initial condition
is a TW of
arbitrary wavenumber $k = 0.295$ and arbitrary amplitude $A_0 = 1.0$ . }
\label{fig2}
\end{figure}

\begin{figure}
\label{fig3}
\caption{Time evolution of $\Phi$ on the $D_{-}$ line. The parameter values are
$a =1, c_{1} = - 1 \mbox{ and } c_{2} = - 1$.  The initial condition is a a
Gaussian
noise of amplitude $\epsilon = 0.01$. The system evolved  towards a TW
attractor
of wavenumber $k = 0.0245$.
}
\end{figure}

\begin{figure}
\caption{Time evolution of $\Phi$ in the non-chaotic region for  $ c_{1} = - 1
\mbox{ and } c_{2} = - 0.8$ . The initial condition is an Eckhaus stable TW of
wavenumber $k = 0.13$ perturbed by random noise of small amplitude  $\epsilon =
0.09$.}
\label{fig4}
\end{figure}

\begin{figure}
\caption{Same as  Fig.\ \protect\ref{fig4}  but for  $c_{1} = - 1 \mbox{ and }
c_{2}
= - 2$ .  The initial condition for $A$ consists of two Eckhaus stable TW of
different wavenumbers  ($k_1 = 0.4 , k_2 = 0.08$) joined  together. The inset
shows
the real part of this initial configuration.
}
\label{fig5}
\end{figure}

\begin{figure}
\caption{ Same as Fig.\ \protect\ref{fig4}  but for  $ c_{1} =  1 \mbox{ and }
c_{2} =
- 0.8$.  The initial condition is a random noise of amplitude $\epsilon =
0.01$.  }
\label{fig6}
\end{figure}

\begin{figure}
\caption{Same as Fig.\ \protect\ref{fig4} but for  $ c_{1} =  - 1 \mbox{ and }
c_{2}
= - 2$ .  The initial condition is an Eckhaus-unstable TW ($k=0.54>k_{E}=0.48$)
slightly perturbed by noise.  }
\label{fig7}
\end{figure}

\begin{figure}
\caption{ The function $\Phi_k \equiv \Phi[A_k]$ as a function of $k$.  The
parameter values are $a =1, c_{1} =   1 \mbox{ and } c_{2} = - 0.8$ . The
values of
$k_{E}$  are indicated by dashed lines. The diamont indicates the point
$\Phi_{k
= 0.64 }$ taken as  initial condition for the simulation in Fig.\
\protect\ref{fig9}   }
\label{fig8}
\end{figure}

\begin{figure}
\caption{Time evolution of $\Phi$ for  $ c_{1} =   1 \mbox{ and }  c_{2} =
-0.8$. The
initial condition is an Eckhaus-unstable TW ($k=0.64>k_{E}=0.48$) slightly
perturbed by noise.
 }
\label{fig9}
\end{figure}

\begin{figure}
\caption{Time evolution of $\Phi$ in the STI region
($c_{1} = 0.0 \mbox{ and } c_{2} = -2$). The initial condition is an
Eckhaus-unstable TW ($k = 0.45 > k_{E} = 0.30$) slightly perturbed by noise.
The
winding number evolution is plotted in the inset.}
\label{fig10}
\end{figure}

\begin{figure}
\caption{Time evolution of $\Phi$ in the Phase Turbulence region  ($ c_{1}=1.75
\mbox{ and } c_{2}= -0.8 $). Solid line: evolution of a  unperturbed unstable
traveling wave. Dotted line: evolution from noise. Dashed line: evolution from
a  slightly perturbed traveling wave. The inset shows final average values   of
$\Phi$ as a function of the $c_{1}$ parameter ($c_{2}=-0.8$). The error bars
indicate the standard  deviation of the fluctuations around the average value.
}
\label{fig11}
\end{figure}

\begin{figure}
\caption{Same as Fig.\ \protect\ref{fig11} but for $ c_{1} =   2 \mbox{ and }
c_{2}
= - 0.8$ .  The initial condition was random noise with an amplitude $\epsilon
=
0.01$, time  step 0.005. In this case 2048 Fourier modes were taken into
account.
Note the  transition occurring arround $t\approx 1000$ to a less  fluctuating
state.  }
\label{fig12}
\end{figure}

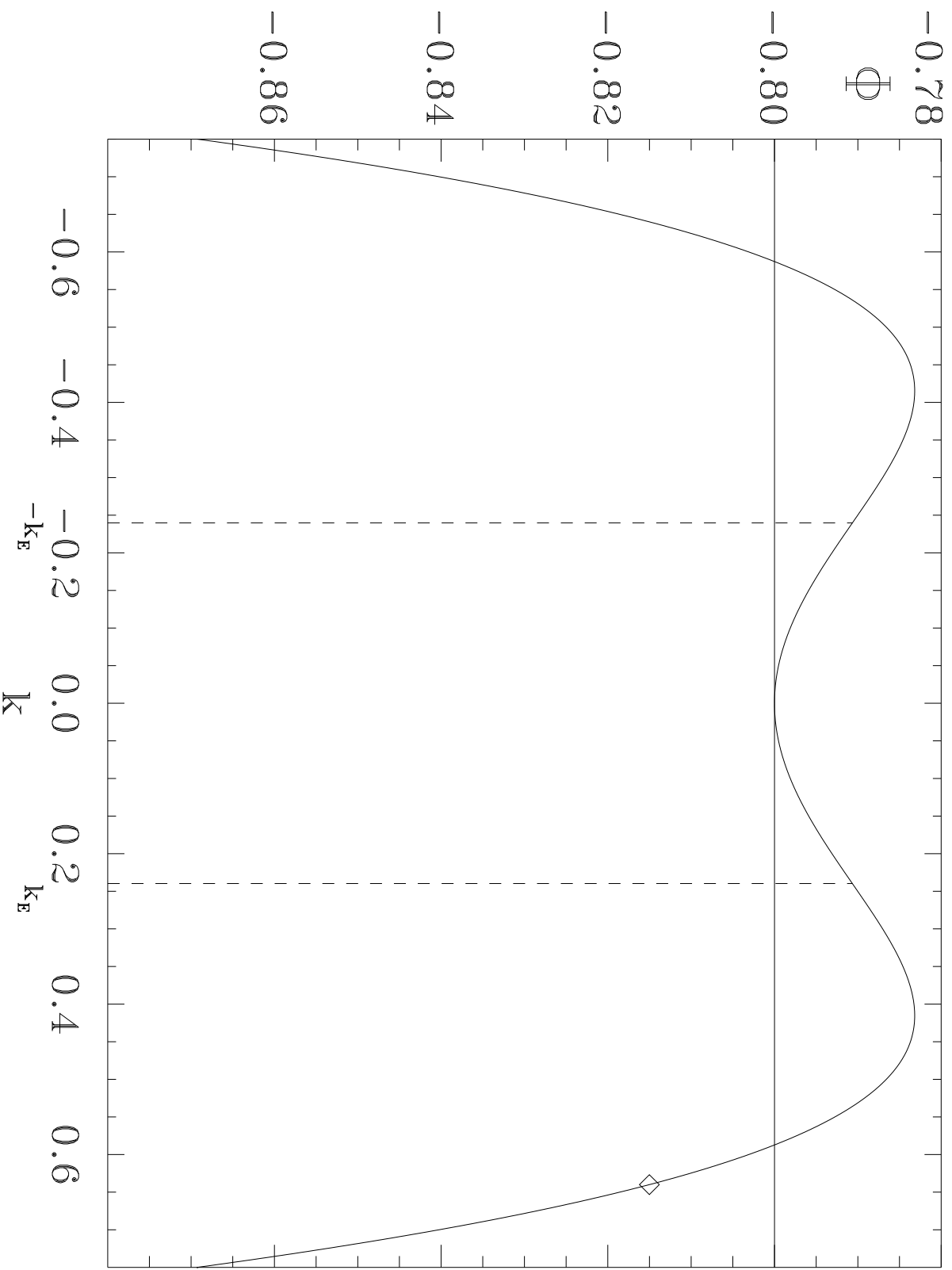
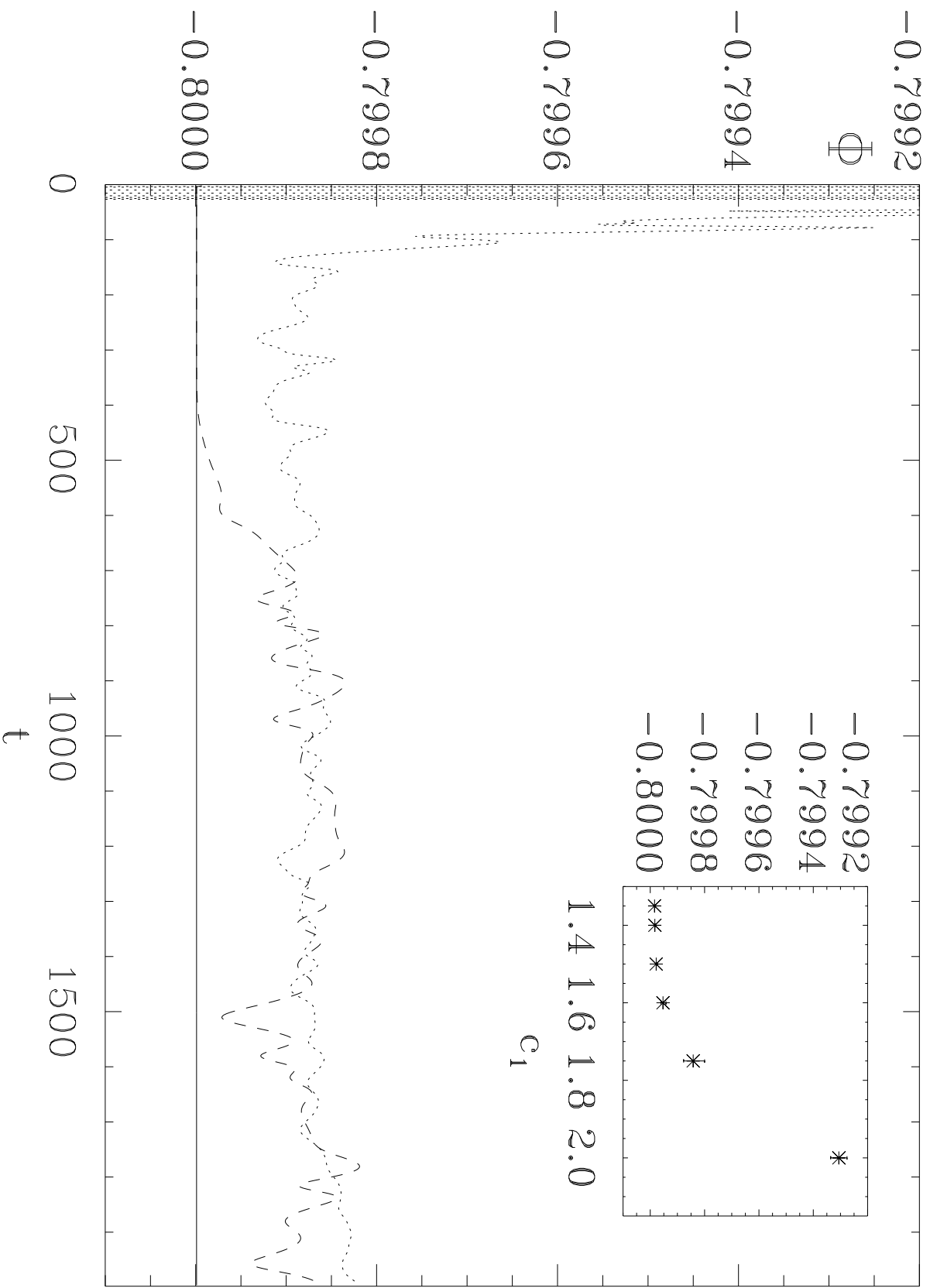
\end{document}